\documentclass[a4paper]{jpconf}
\usepackage{graphics,graphicx,dcolumn,bm,fleqn,epic,eepic,float,epsfig}
\usepackage{amssymb,amsmath,multirow,rotate,color,float}
\usepackage{epstopdf}
\usepackage{times}
\usepackage{color}
\usepackage{soul}                    
\usepackage{ulem, caption, float}

\definecolor{red}{rgb}{1,0,0}
\definecolor{green}{rgb}{0,1,0}
\definecolor{blue}{rgb}{0,0,1}

\begin{document}

\title{Principal wind turbines for a conditional portfolio approach 
to wind farms}

\author{Vitor V.~Lopes$^{1,2}$, 
        Teresa Scholz$^{3,4}$, 
        Frank Raischel$^5$,         
        Pedro G.~Lind$^6$}

\address{$^1$DEIO-CIO, Science faculty, University of Lisbon, Portugal\\
         $^2$Universidad de las Fuerzas Armadas-ESPE, Latacunga, Ecuador\\         
         $^3$Center for Theoretical and Computational Physics, University  
             of Lisbon, Portugal\\ 
         $^4$Energy Analysis and Networks Unit, National Laboratory of 
             Energy and Geology, Lisbon, Portugal\\
         $^5$Center for Geophysics, IDL, University of Lisbon, Portugal\\
         $^6$ForWind-Center for Wind Energy Research, Institute of Physics, 
             Carl-von-Ossietzky University of Oldenburg, Oldenburg, Germany}

\date{\today}

\ead{teresa.scholz@lneg.pt}

\begin{abstract}
    We introduce a measure for estimating the best risk-return relation of power production in wind farms within a given time-lag, conditioned to the velocity field. The velocity field is represented by a scalar that weighs the influence of the velocity at each wind turbine at present and previous time-steps for the present ``state'' of the wind field. The scalar measure introduced is a linear combination of the few turbines, that most influence the overall power production. This quantity is then used as the condition for computing a conditional expected return and corresponding risk associated to the future total power output.
\end{abstract}

\section{Introduction}
\label{section:Introduction}

    Wind energy is becoming a top contributor to the renewable energy mix due to rather high capacities and generation costs that are becoming competitive with conventional energy sources\cite{Wen2009}. However, wind energy systems suffer from a major drawback, the fluctuating nature of their source\cite{prlpatrick}, which imposes a challenge to the wind power producer when it comes to trading on the liberalized electricity markets. For that, participants must bid in advance and the uncertainty of wind power production can lead to differences between the committed and actually produced energy. This imbalance may result in the payment of penalties which decreases the revenue. Therefore, to achieve maximum profit, it is necessary to develop optimal offering strategies. A review on the methods employed for deriving bidding strategies is given by\cite{Li2011}.

    In this study we propose a measure to asses the return of a wind farm in terms of risk that can be used to develop bidding strategies for market participation. Our measure is based on the mean-variance portfolio (MVP) or risk-return approach first introduced in the context of portfolio selection by Markowitz\cite{Markowitz1952}. Markowitz defines the return on a portfolio as a weighted sum of random variables where the investor can choose the weights. For investment he proposes to assess the ratio between the expected return and the associated risk, which he defines as the standard deviation of return. An investment decision then is a trade-off between risk and expected return, since ``the portfolio with maximum expected return is not necessarily the one with minimum variance''\cite{Markowitz1952}. However, for a given amount of risk, MVP allows to maximize the expected return for a given level of risk or equivalently to minimize the risk for a given level of expected return.

    In the context of wind energy this translates into optimizing the trade-off between maximizing wind power output and minimizing its variability. MVP has been employed in the framework of wind energy by Roques et al\cite{Roques2010} to define optimal cross-countries wind power portfolios. Using historical wind production data from five European countries the authors attempt two case studies. First, they optimize the wind power output and, second, maximize the wind power contribution to system reliability. The risk-return approach has also been employed by Kitzing\cite{Kitzing2014} with the purpose to assess the risk implications of two support instruments, feed-in tariffs and feed-in premiums.

    This paper introduces a new approach of risk-return evaluation by including a dependence on the state of the full wind farm. More precisely, our risk-return approach is {\it conditioned} to the wind field at each time-step. It is known that such conditional stochastic approaches allow to tackle the non-stationary character of wind\cite{prlpatrick,lind}.

    The paper is organized as follows. Section \ref{section:Data} describes the used datasets. In Sec.~\ref{section:Methodology} the employed methodology is introduced, first, the conditional risk-return quotient based on this scalar is defined and then a scalar observable is derived for quantifying the velocity field taken at specific time instants. In Sec.~\ref{section:Results} we present and explain the results obtained for a wind farm in Portugal and Sec.~\ref{section:Conclusions} concludes the paper.

\section{Data: The wind field and total power production}
\label{section:Data}

    The data analyzed in this manuscript were obtained from a wind park with $80$ turbines located in a mountainous region of Portugal. The data comprise three years of measurements with a ten-minute sampling time ($1.7\times 10^{-3}$ Hz). They consist of two sets, the wind velocity field observed at each wind energy converter (WEC) in the farm and the farm's total power production. 

    The velocity field is represented as a matrix $\mathbf{V} \in \mathbb{C}^{N_W\times N_T}$, where each entry $V_{n}(t)=V^{(x)}_{n}(t)+iV^{(y)}_{n}(t)$ corresponds to the velocity vector with $x$- and $y$-components $V^{(x)}_{n}(t)$ and $V^{(y)}_{n}(t)$ at the WEC labelled as $n=1,\dots,N_W$ and time $t=1,\dots,N_T$:

    \begin{equation}
        \mathbf{V} = \left [
        \begin{array}{ccccc} 
            V_{1}(1)   & V_{1}(2)   & V_{1}(3)   & \dots & V_{1}(N_T) \\ 
            V_{2}(1)   & V_{2}(2)   & V_{2}(3)   & \dots & V_{2}(N_T) \\ 
            V_{3}(1)   & V_{3}(2)   & V_{3}(3)   & \dots & V_{3}(N_T) \\ 
            \vdots    & \vdots     & \vdots    & \ddots & \vdots \\
            V_{N_W}(1) & V_{N_W}(2)  & V_{N_W}(3) & \dots & V_{N_W}(N_T) 
        \end{array}
        \right ] ,
    \end{equation} 

    where $V_i(t)$ is the velocity at turbine $i$ at time-step $t$. Vertical velocities are always neglected.    

    Our aim is to estimate the expected power production $P(t+\tau)$ at a time-lag $\tau$ after the present time $t$, using the velocity field observations at time $t$ as input. The methodology is described in the next section.

\section{Methodology and data analysis}
\label{section:Methodology}

    This section describes the methodology used for the data analysis. A new conditional risk-return approach to power production in a wind farm is introduced. It is stressed that any risk measure should take into account the present ``state'' of the wind farm. The straightforward way to define this state is through the full set of wind velocities at a particular time. However, as we stress below, the full wind field would imply a large amount of data to determine the state of the system. Therefore, instead of the full velocity field, we define a scalar quantity that is also capable to characterize the state of a spatially extended system such as a wind farm.

    The usual way to define the percentage power return $r(t)$ is\cite{Kitzing2014}

    \begin{equation}
        r(t) = \frac{P(t+\Delta t)-P(t)}{P(t)},
        \label{eq1}
    \end{equation}
    
    where $\Delta t$ is a fixed time-lag.
    Having the time series of the returns we can then define the expected power return by

    \begin{equation}
        \hat{r} = \int_{-\infty}^{\infty} r \rho(r(t))dr ,
        \label{expectedPstand}
    \end{equation}

    and the associated risk is given by the variance of the expected returns, namely

    \begin{equation}
        \Delta r = \int_{-\infty}^{\infty} (r-\hat{r})^2 \rho(r(t))dr ,
        \label{deviationPstand}
    \end{equation}

    where $\rho(r(t))$ is the probability for having a return $r(t)$ at time $t$. 

    In both Eqs.~(\ref{expectedPstand}) and (\ref{deviationPstand}) the return $r$ considers the total power output in the wind farm. Alternative choices are possible, e.g.~to account only for the power of a subset of all WECs, the ones that represent the most the power output in the wind farm. An improvement to the power sum of the total or a subset of WECs would be a weighted sum, but in all cases it would yield an expected value and a variance (risk) independent of the present wind velocity field.

    We propose instead to consider a proper quantity, ``quantifying''  the wind velocity field, and use it as condition for computing the expected return and corresponding variance. This state of one wind farm could be defined as the set of values of the wind velocities at instant $t$. However, since wind farms typically contain $N_W\sim 100$ wind turbines, each one with $N_s\sim 50$ admissible wind speed states after proper binning\cite{Lopes2012,scholz2014}, the phase space for the wind farm would comprehend approximately $50^{100}$ possible states.

    To overcome this shortcoming, we choose a weighted sum based on the principal component analysis (PCA) of the wind speed at all the turbines and several time-lags. This weighted sum is truncated at a given order $q$, and we symbolize it henceforth as $S^{(q)}(t)$ (defined in Eq.~(\ref{eq:S_corr})), which we introduce as follows.

    The PCA is performed through the eigenvalue decomposition of a matrix constructed from the covariance matrix of wind velocities. Given a set of time-lags $\mathcal{T} = \{\tau_1, ..., \tau_{N_\tau}\}$, for each pair of turbines $i$ and $j$, we compute the covariance associated to their wind velocities referenced to given time-lags $\tau_k, \tau_l \in \mathcal{T}$ by
    
    \begin{align}
        \mathbf{M}_{(i\tau_k)(j\tau_l)} &= \{ C_{ij(\tau_l - \tau_k)} \}_{i,j = 1,\dots,N_W} \cr  
            &=
        \{ \langle (V_{i}(t + \tau_k)-\langle V_{i}(t + \tau_k)\rangle)
                (V_{j}(t + \tau_l)-\langle V_{j}(t + \tau_l)\rangle)^\ast 
        \rangle \}_{i,j = 1,\dots,N_W} \cr
                &= 
        \left [
        \begin{array}{ccccc} 
            M_{(1\tau_k)(1\tau_l)}  & M_{(1\tau_k)(2\tau_l)} & \dots & M_{(1\tau_k)(N_W\tau_l)} \\
            M_{(2\tau_k)(1\tau_l)}  & M_{(2\tau_k)(2\tau_l)} & \dots & M_{(2\tau_k)(N_W\tau_l)} \\
            \vdots & \vdots & \ddots & \vdots \\
            M_{(N_W\tau_k)(1\tau_l)}  & M_{(N_W\tau_k)(2\tau_l)} & \dots & M_{(N_W\tau_k)(N_W\tau_l)} \\
        \end{array}
        \right ] ,
    \end{align}

    where $i,j=1,\dots,N_W$, $l,k = 1,\dots,N_{\tau}$ and $\langle  \rangle$ denotes the average over time $t$ and $^{\ast}$ denotes the complex conjugate. 

    Taking $\tau_{\ell}, \tau_k=1,\dots,N_{\tau}$ the covariances can be embedded in a master matrix defined as 

    \begin{eqnarray}
    \mathbf{M} &=& \left [
        \begin{array}{cccc} 
            \mathbf{M}_{(i0)(j0)} & \mathbf{M}_{(i0)(j1)} & \dots & \mathbf{M}_{(i0)(jN_{\tau})}\\
            \mathbf{M}_{(i1)(j0)} & \mathbf{M}_{(i1)(j1)} & \dots & \mathbf{M}_{(i1)(jN_{\tau})}\\
            \vdots & \vdots & \ddots & \vdots \\
            \mathbf{M}_{(iN_{\tau})(j0)} & \mathbf{M}_{(iN_{\tau})(j1)} & \dots & 
                \mathbf{M}_{(iN_{\tau})(jN_{\tau})} \\  
        \end{array}
        \right ]   
        \label{mastermatrix}
    \end{eqnarray}

    which is a symmetric matrix of dimension $N_WN_{\tau}\times N_WN_{\tau}$. 
        
    The master-matrix has two important properties. The first one is the symmetry $M_{(i\tau_k)(j\tau_l)} = M_{(j\tau_l)(i\tau_k)}$. Notice that $\langle V_{i}(t + \tau_k)\rangle = \frac{1}{N_T-\tau_k} \sum_{t=\tau_k+1}^{N_T} V_{i}(t)$. Therefore, for sufficiently large number of measures, $N\gg \tau_i$, $\forall i$, we take $\langle V_{i}(t + \tau_k)\rangle = \langle V_{i}(t) \rangle$, $\forall i$. Thus, $\mathbf{M}_{(i\tau_k)(j\tau_l)} = \mathbf{M}_{(i0)(j(\tau_l-\tau_k))} = \mathbf{M}_{(j0)(i(\tau_k-\tau_l))} = \mathbf{M}^T_{(i0)(j(\tau_k-\tau_l))} = \mathbf{M}^T_{(i\tau_l)(j\tau_k)}$. The master-matrix $\mathbf{M}$ is therefore symmetric. Notice however that it can have complex eigenvalues and eigenvectors since $V_i(t+\tau_k)$ is complex $\forall i,k$. The second property is that, assuming approximation $\langle V_{i}(t + \tau_k)\rangle = \langle V_{i}(t) \rangle$, the master-matrix is Toeplitz by blocks.
    
    \begin{figure}[t]
        \centering
        \includegraphics[width = 0.4\textwidth]{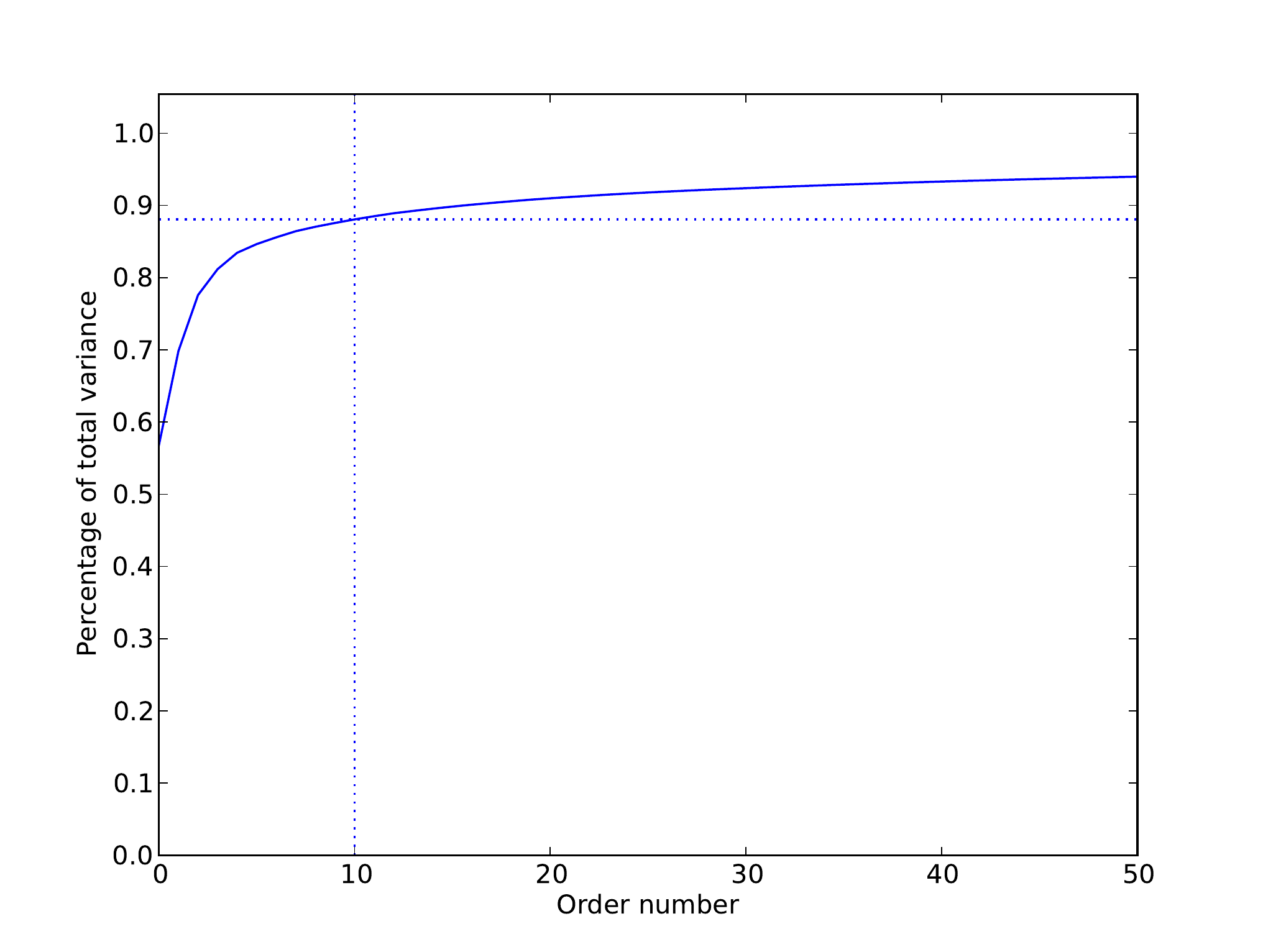}%
        \includegraphics[width = 0.4\textwidth]{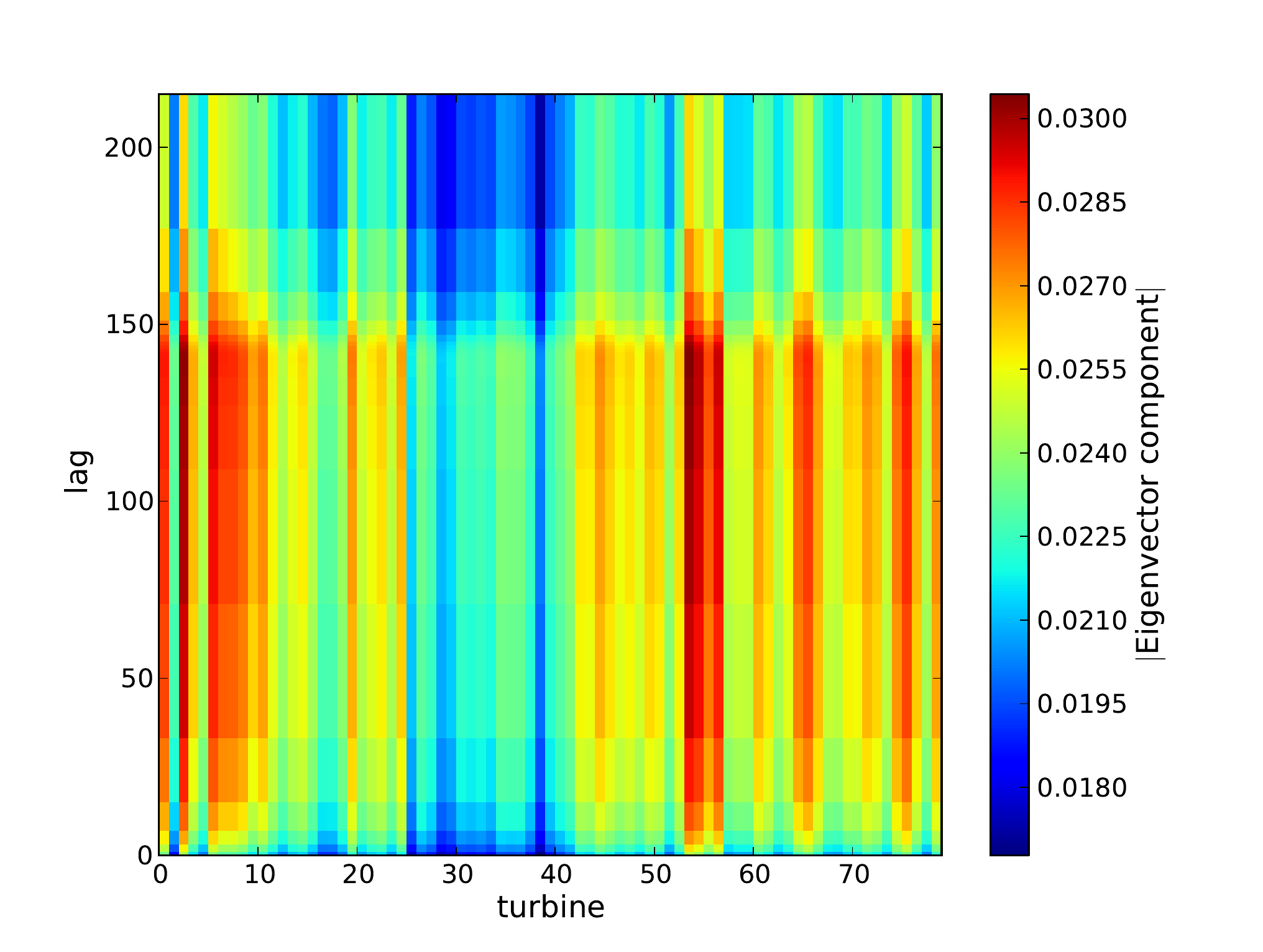}%
        \caption{Percentage of total variance expressed by the first $50$ 
                eigenvectors (left) and the absolute value of the 
                components of the eigenvector corresponding to the first 
                eigenvalue (right) of the master matrix $M$.}
        \label{fig1}
    \end{figure}
        
    The diagonalization of this covariance matrix allows to compute principal directions which form the set of components uncorrelated with each other. These directions in phase space are defined by linear combinations of the wind speed at each single turbine and different time-lags. The principal directions corresponding to the largest eigenvalues are the ones along which the system fluctuates strongly. Therefore, selecting from these linear combinations the few ones corresponding to the largest eigenvalues gives a subset of the most ``influent'' turbines for the global wind velocity state of the wind farm.
        
    Figure~\ref{fig1} on the left shows the percentage of the total variance that is expressed by the first $m$ (largest) eigenvalues of the master matrix. It can be seen that the ten largest eigenvalues already account for almost $90\%$ of the total variance in the data (check dotted lines in Fig.~\ref{fig1}). These facts will be important below to interpret our results.

    Further the principal direction corresponding to the largest eigenvalue alone comprehends already almost $50\%$ of signal's variance. Its components are shown on the right of Fig.~\ref{fig1}. It can be concluded that the contribution of the turbines is not equal throughout the park, for instance turbines $54$ to $57$ have a high influence, turbines $25$ to $42$ a comparatively low one. Considering the contributions of the time-lags, i.e.~the influence of information from the past, Fig.~\ref{fig1} shows that the contribution starts out small for zero time lag and quickly reaches a plateau. It then decays rapidly for lags larger than one day. This pattern is especially prominent for turbines with lower contributions.
    
    \begin{figure}[htb]
        \centering
        \includegraphics[width = 0.98\textwidth]{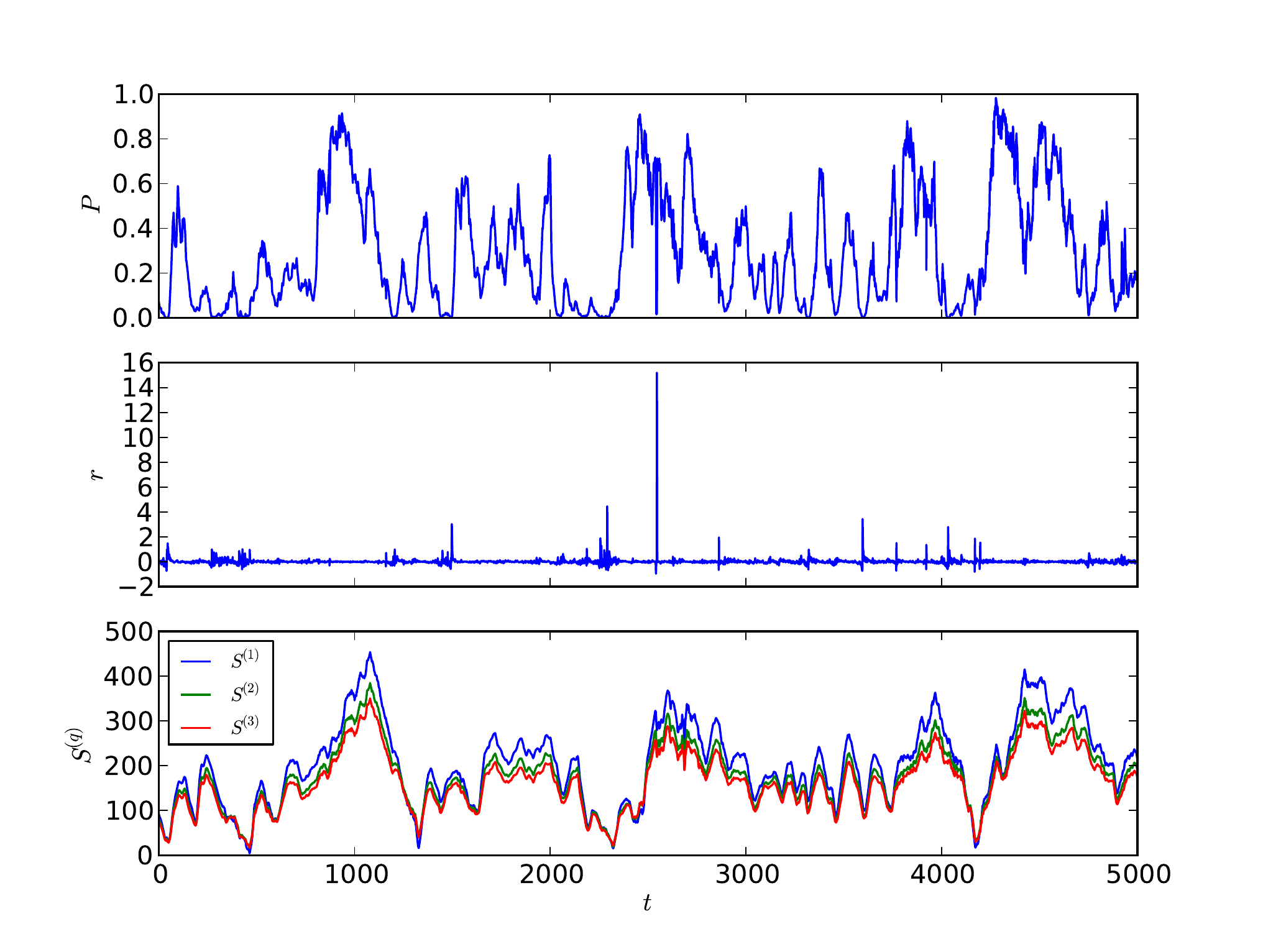}
        \caption{Subset of the time series of the wind park's total power 
                output (top), return (middle) and wind park state $S^{(q)}$ 
                for $q={1,2,3}$ (bottom).}
        \label{fig2}
    \end{figure}
    \begin{figure}[htb]
        \centering
        \includegraphics[width=0.49\textwidth]{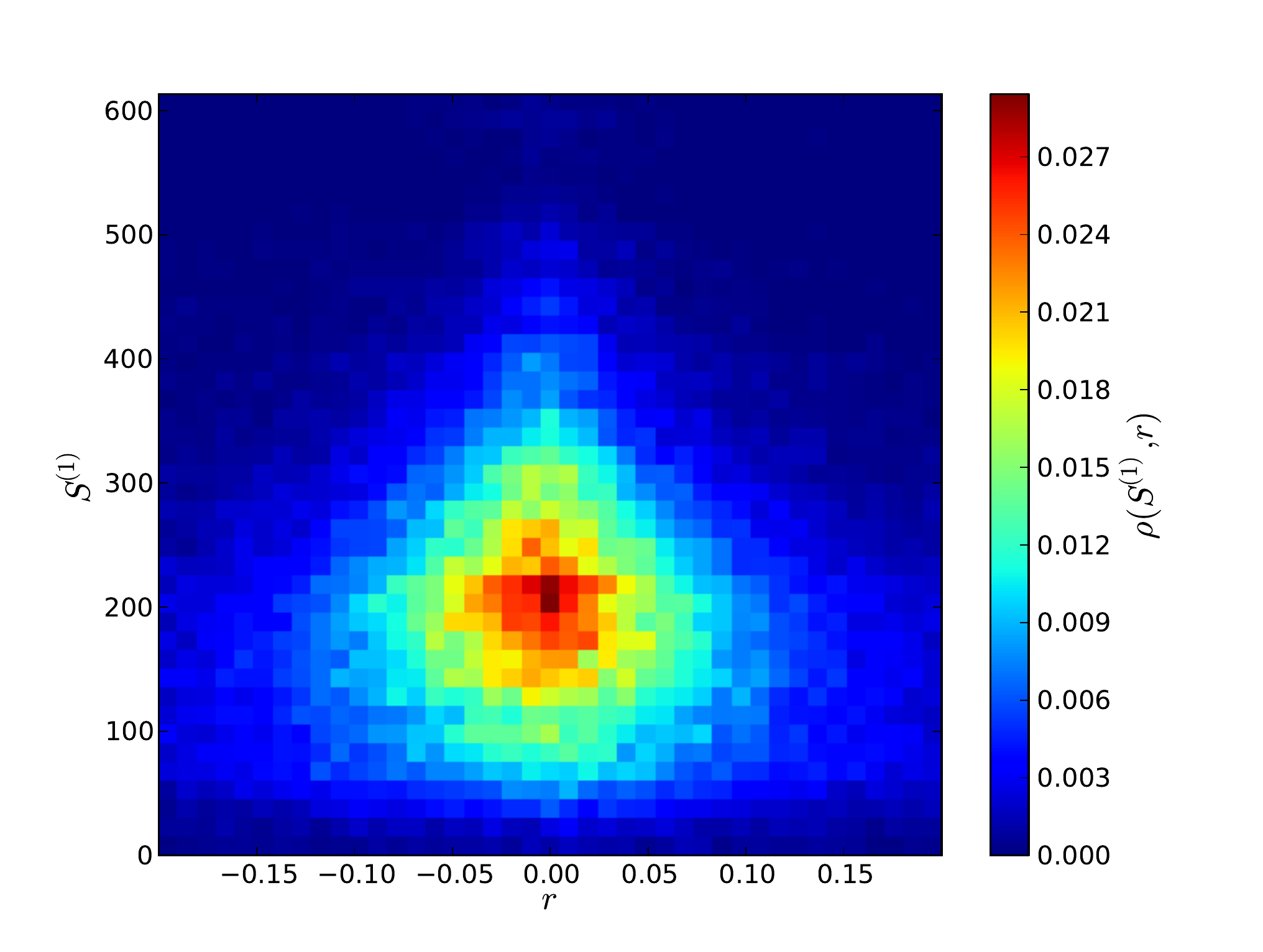}%
        \includegraphics[width=0.49\textwidth]{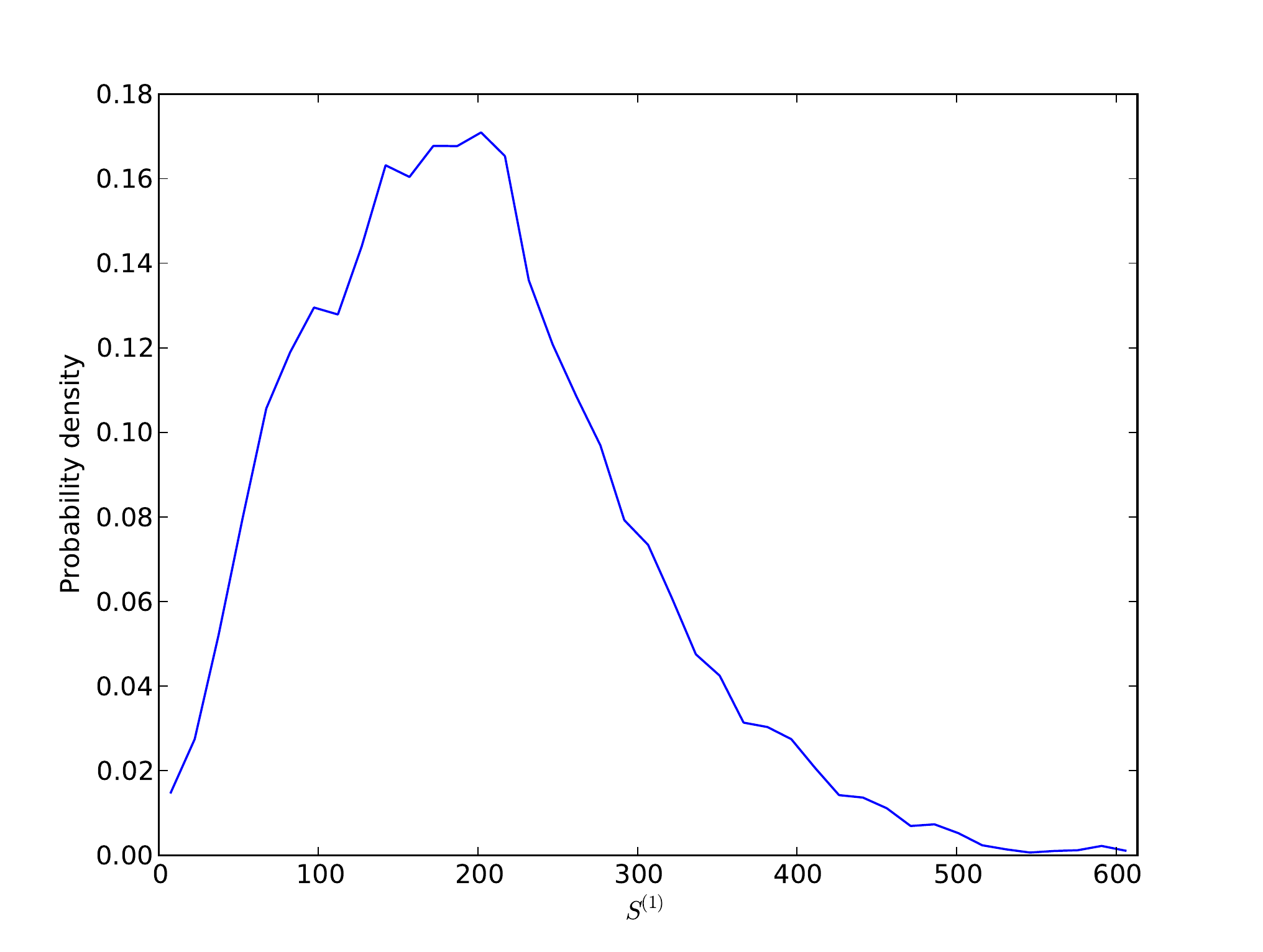}%
        \caption{Joint probability density function of the wind park 
        state $S^{(1)}$ and the power return $r$ (left)
        and probability density function of the wind park state $S^{(1)}$ 
        (right).}
        \label{fig3}
    \end{figure}   
    
    Next, we use the insight provided by a set of eigenvectors ordered according to the magnitude of their corresponding eigenvalue to define a scalar quantity for characterizing the state of the wind farm. Namely, we consider the first $q$ eigenvalues $\lambda_1, \dots, \lambda_q$ of $\mathbf{M}$ with $q = 1,\dots,N_W N_{\tau}$, together with their corresponding eigenvectors and define the scalar quantity 
    
    \begin{equation}
        S^{(q)}(t) =\frac{\sum_{i=1}^q \left |\lambda_i \sum_{j=1}^{N_W} 
                \sum_{k=1}^{N_{\tau}} 
                \omega_{ij}V_j(t-\tau_k) \right \vert}{\sum_{i=1}^q \left|\lambda_i  
                \right|},
        \label{eq:S_corr}
    \end{equation}

    where $\omega_{ij}$ describes the $j$-th component of the eigenvector to the $i$-th eigenvalue $\lambda_i$ and $\tau_k$ is given in units of the time increment between successive measurements of the wind speed.
    \begin{figure}[htb]
        \centering
        \includegraphics[width=0.95\textwidth]{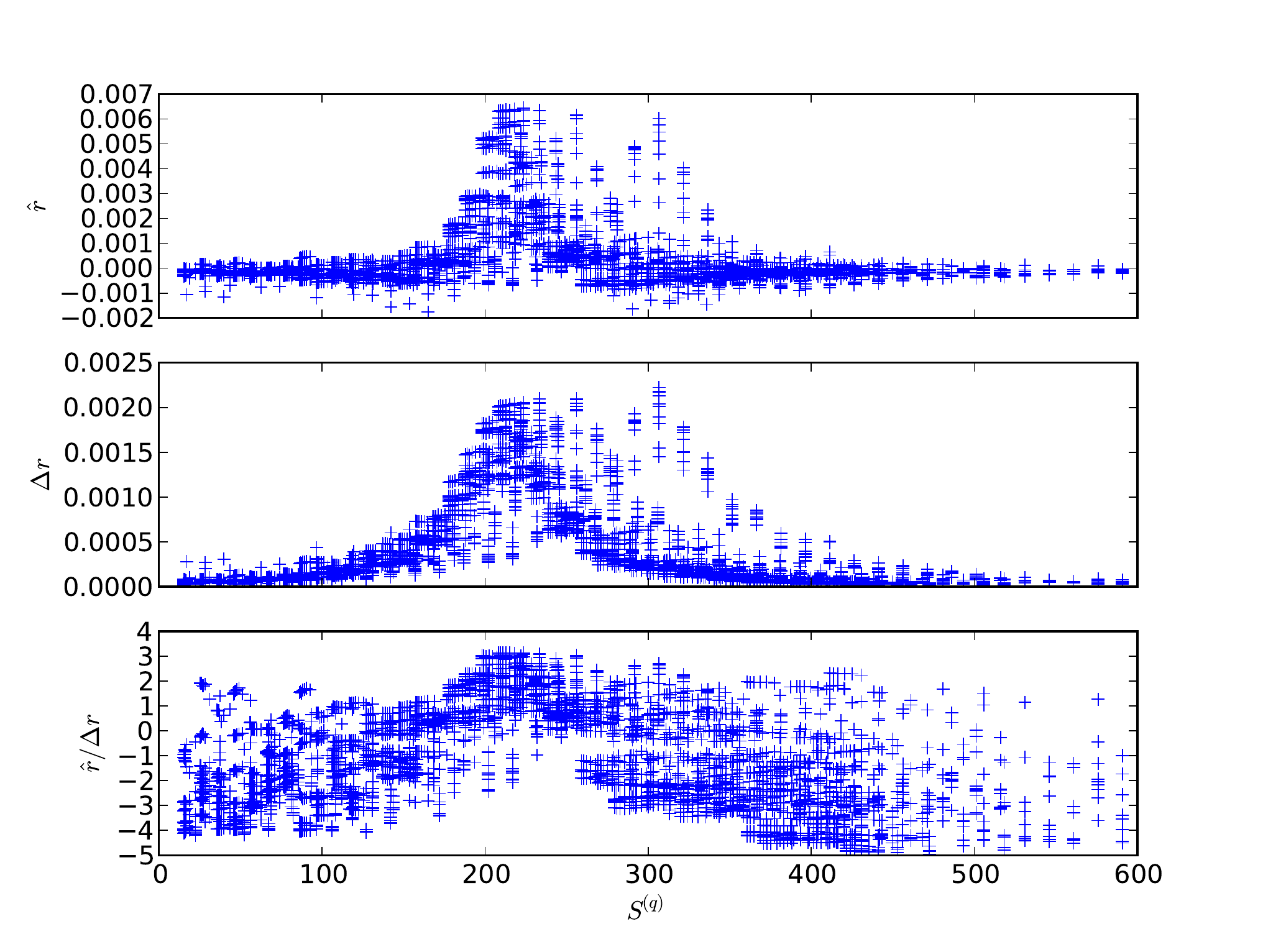}
        \caption{Conditional returns $\hat{r}(q,\tau,S^{(q)}(t))$ 
                (top), risks $\Delta r (q,\tau,S^{(q)}(t))$ 
                (middle) and risk-return quotient 
            $\hat{r}(q,\tau,S^{(q)}(t))/\Delta r (q,\tau,S^{(q)}(t))$ 
                (bottom) as a function of the state $S^{(q)}$ for 
                several values of $q=1,\dots,10$ and $\tau=1,\dots,10$.} 
        \label{fig4}
    \end{figure}

    We argue here that $S^{(q)}(t)$ quantifies the wind velocity state of the wind farm at time $t$.
    Since the eigenvalues are ordered the first sum in Eq.~(\ref{eq:S_corr}) comprehends the $q$ largest eigenvalues. The other two sums are in the set of WECs and in (previous) times. Thus, for each choice of WEC $j$ and time-delay $\tau_k$ the quantity $S$ uses a weight given by the $j$-th component of the $i$-th most influent eigenvector. In other words, by weighting a sum of velocity measurements across the wind farm and, simultaneously, at different time-steps, the quantity $S^{(q)}$ incorporates both spatial and temporal information of the wind speed observed at the wind farm. We call $S^{(q)}$ the state of the wind farm.

    Figure~\ref{fig2} (bottom) shows the evolution of $S^{(q)}$ for $q=1,2$ and $3$ together with the time series of the wind park's total power output (top) and the corresponding returns given by Eq.~(\ref{eq1}) (middle). One can observe that adding information from the third largest eigenvalue and its corresponding eigenvector leads to very small changes in $S^{(q)}$, especially when $S^{(q)}$ is small. Similarly to what was already concluded when addressing Fig.~\ref{fig1}, such similar evolutions indicate that the lowest $q$ values already capture most of the information in the velocity field, at least relatively to its energy content.

    Using $S^{(q)}$, we introduce a {\it conditional} risk-return approach, where expected return and risk are given respectively by
    \begin{equation}
        \hat{r}(q,\tau,S^{(q)}(t)) = \int_{-\infty}^{\infty} r 
            \rho(r(t+\tau)\vert S^{(q)}(t))dr
        \label{expectedP}
    \end{equation}
    and
    \begin{equation}
        \Delta r(q,\tau,S^{(q)}(t)) = 
        \int_{-\infty}^{\infty} (r-\hat{r})^2 \rho(r(t+\tau)\vert S^{(q)}(t))dr .
        \label{deviationP}
    \end{equation}

    The joint probability density function $\rho(r(t),S^{(1)})$ for both the state $S^{(1)}(t)$ and the total power output return $r(t)$ is shown in Fig.~\ref{fig3} (left) together with the marginal probability density function for $S^{(1)}$ (right). The joint distribution is approximately symmetric around $r=0$ and its variance decreases with the value of $S$, i.e.~the expected variability of the power decreases with the higher intensity of the wind field.

    Both density functions in Fig.~\ref{fig3} are needed for deriving the conditional probability $\rho(r(t+\tau)\vert S^{(q)}(t))= \rho(r(t+\tau),S^{(q)}(t))/\rho(S^{(q)}(t))$ needed for the computation of the conditional risk-return measure in Eqs.~(\ref{expectedP}) and (\ref{deviationP}). 
    \begin{figure}[htb]
        \centering
        \includegraphics[width=0.98\textwidth]{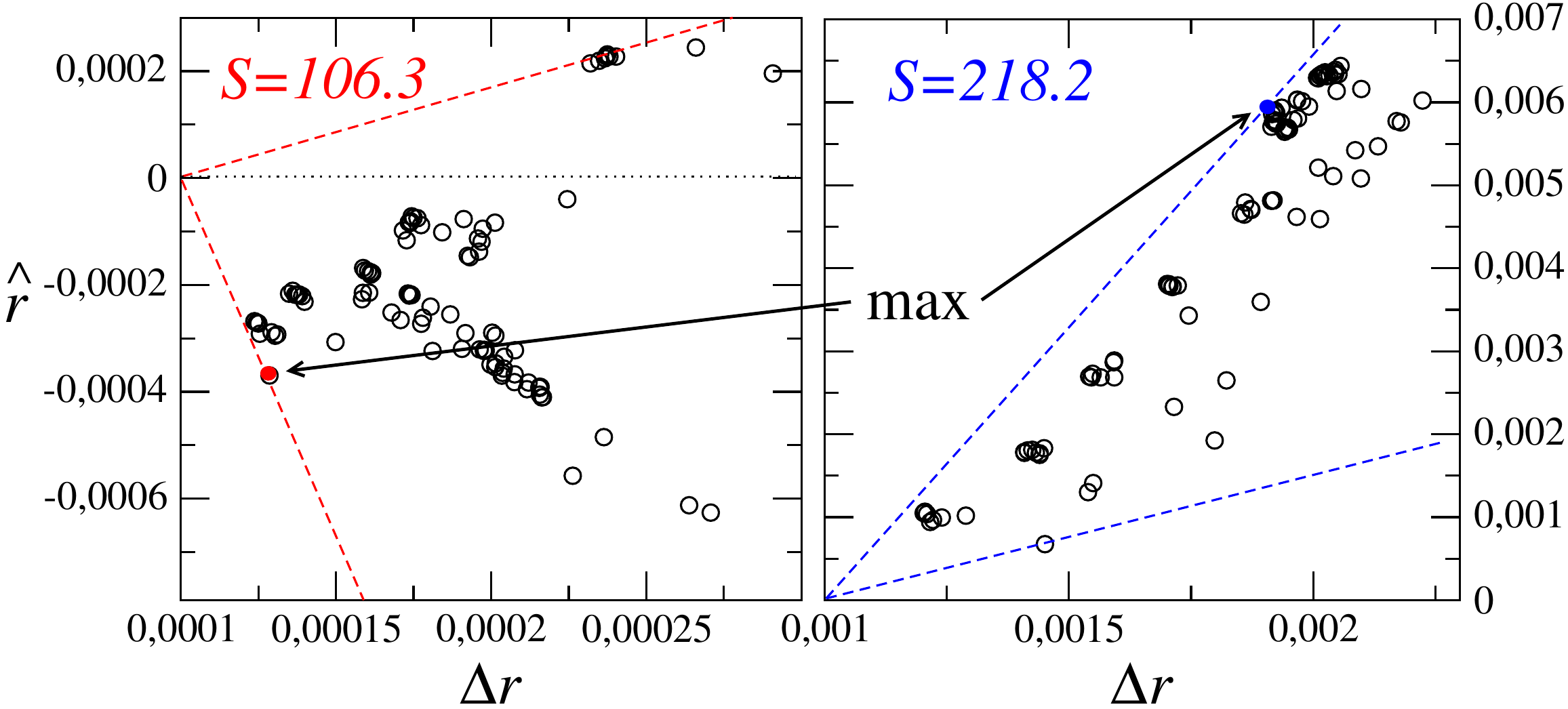}
        \caption{Risk-return diagrams for two different values of 
                $S^{(q)}(t)$. Each circle corresponds to a pair ($q$, $\tau$). For  $S^{(q)}(t)=106.3$ the maximum of 
                the risk-return quotient is obtained for $q_{max}=1$ 
                and $\tau_{max}=1$ (red bullet), for $S^{(q)}=218.2$ 
                however, one obtains the values $q_{max}=10$ and 
                $\tau_{max}=4$ (blue bullet).} 
        \label{fig5}
    \end{figure}
    \begin{figure}[htb]
        \centering
        \includegraphics[width=0.98\textwidth]{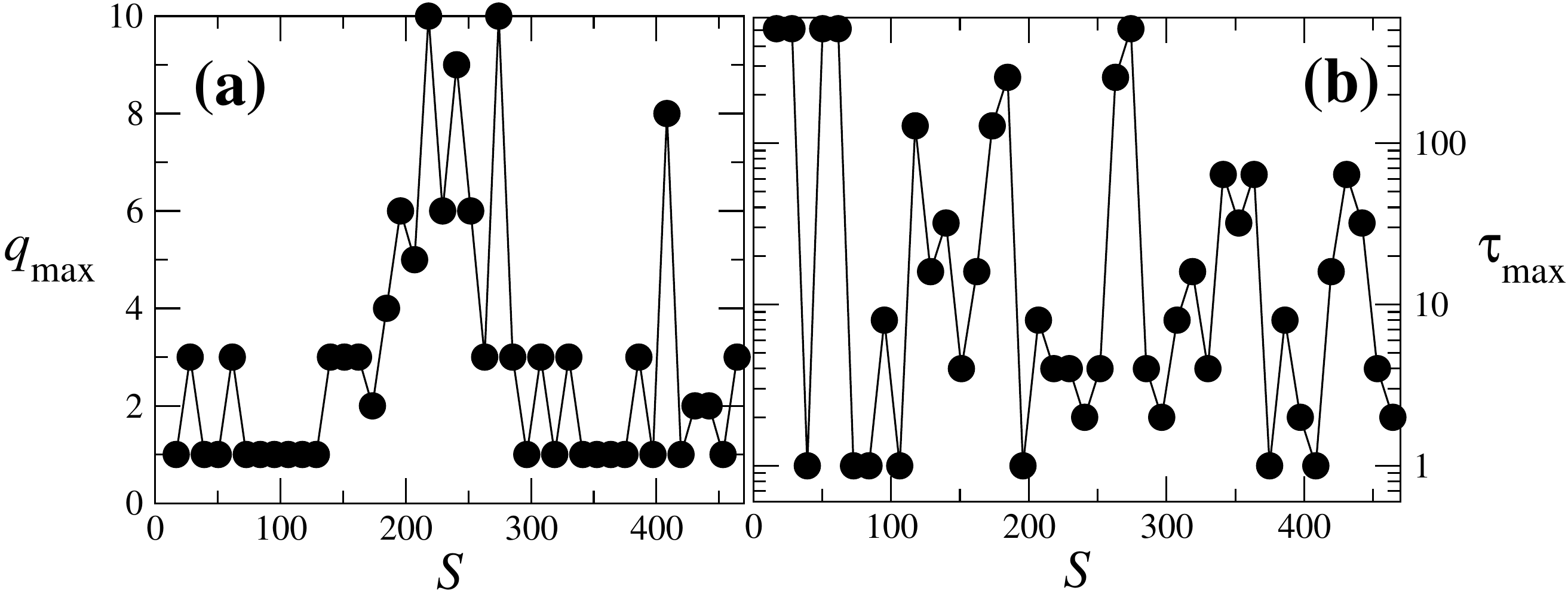}
        \caption{Values that should be selected for $q_{max}$ and 
                $\tau_{max}$ for each state of the wind farm.} 
        \label{fig:fig7}
    \end{figure}

\section{Results}
\label{section:Results}

    Using Eqs.~(\ref{expectedP}) and (\ref{deviationP}), we can now compute the conditional returns $\hat{r}(q,\tau,S^{(q)}(t))$, risks $\Delta r(q,\tau,S^{(q)}(t))$ and risk-return ratios $\hat{r}/\Delta r$ as a function of the state $S^{q}(t)$, i.e.~fixing the values of $q$ and $\tau$. Results are shown in Fig.~\ref{fig4}, where $q$ and $r$ range between $1$ and $10$. Important to notice here is that the conditional risk-return ratio varies a lot with $q$ and $\tau$ within a wind farm state $S^{(q)}(t)$. Therefore, a possible optimal bidding strategy would be to take into account which values of $q$ and $\tau$ yield the maximum ratio. 

    This sensitivity of the optimal risk-return ration to the state of the wind farm is exemplified in detail in Fig.~\ref{fig5} for two particular values of $S^{(q)}(t)$: for  $S^{(q)}=106.3$ the maximum of the risk-return quotient is obtained for $q_{max}=1$ and $\tau_{max}=1$ (red bullet), for $S^{(q)}=218.2$ however, one obtains the values $q_{max}=10$ and $\tau_{max}=4$ (blue bullet).

    Figure~\ref{fig:fig7} shows the time series of optimal values, $q_{max}$ and $\tau_{max}$ for the full range of possible states observed in the wind farm analyzed in this study.

    The values of $q_{max}$ have values ranging from one up to ten. The lowest values - typically $q_{max}=1,2$ or $3$ - are attained for very weak winds (small values of $S$, left side of the spectrum) or for wind gusts (large values of $S$, right side of the spectrum). This can be easily explained: the two extreme situations are the ones for which the wind turbines are more synchronized and therefore most of the variance is already included in very first eigen-modes, corresponding to the lowest values of $q_{max}$. In the middle range of the wind speed (and also of the state $S$) the wind farm is far more heterogeneous and therefore higher order eigen-modes are necessary to explain the variability of the power output, yielding larger values for $q_{max}$.

    As for $\tau_{max}$ the variability is much stronger, ranging from one up to several hundreds. The values chosen for $\tau_{max}$ were the ones that numerically maximize the absolute value of the risk-return ratio. Whether for other values of $\tau$ the risk-return ratio for a given state $S$ is {\it approximately} the same needs to be still investigated.

\section{Conclusions and discussion}
\label{section:Conclusions}

    In this paper we extend the standard portfolio analysis introducing a maximum risk-return ratio conditioned to the present state of the system and apply it to one wind farm, by taking the observed wind velocity field into account. For each defined wind farm state, the corresponding optimal risk-return ratio yields a particular time-lag ($\tau$) which gives the best time-horizon to make power output forecasts.

    Our results provide evidence that it is sensible to select the prediction horizon depending on the present state of the wind speed field instead of fixing it, as it is standardly done. Since, as we know, fixing the time-lag for the forecast, independently of the present state of the wind field, can lead to the underestimation of the risk levels for a given expected return or to overestimation of the return level for a given expected risk.

    We conjecture that this dependence on the state of the system happens due to the non-stationary and intermittent character of the wind velocity field\cite{prlpatrick,lind}, not only at one WEC but at the level of  the entire wind park. Therefore, such a procedure can be helpful in other situations dealing with non-stationary and intermittent sets of data, such as the ones commonly observed in brain research and finance.
    
    Two important remarks raise from the conclusions of our results. First, the conditional portfolio approach can be taken to derive other more sophisticated risk measurements, such as the value-at-risk for the total power output\cite{Riskmetrics}. 

    Second, the time evolution of the state of the wind farm as defined above, while properly reflecting the physical situation of interest here, could be studied in deeper detail elsewhere, particularly in what concerns its (non)-stationary character and possible intermittency of its increments in different time-scales. Depending on the outcome a Langevin approach already successfully applied to power output and other WEC properties\cite{prlpatrick,pre}, may provide further insight of the evolution of wind farms. These and other issues will be addressed elsewhere.

\section*{Acknowledgments}

    The authors thank Philip Rin, Jo\~ao P.~da Cruz for useful discussions and GENERG, SA. for providing the original data. The authors thank Funda\c{c}\~ao para a Ci\^encia e a Tecnologia for financial support under PEst-OE/FIS/UI0618/2011, PEst-OE/MAT/UI0152/2011, FCOMP-01-0124-FEDER-016080, SFRH/BPD/65427/2009 (FR) and SFRH/BD/86934/2012 (TS). PGL thanks the German Environment Ministry for financial support (0325577B).


\end{document}